\documentclass[11pt,twoside]{article}
\usepackage{./asp2014}

\newcommand{\Msun}{M$_\odot$}

\usepackage{draftwatermark}
\SetWatermarkText{DRAFT}
\SetWatermarkScale{1}

\aspSuppressVolSlug
\resetcounters

\begin{document}

\title{A six-dimensional tomographic view of Galactic star-formation}
\author{Laurent Loinard and Mark J.\ Reid \\
\affil{Centro de Radioastronom\'{\i}a y Astrof\'{\i}sica, Universidad Nacional Aut\'onoma de M\'exico, Morelia, Michoac\'an, M\'exico \email{l.loinard@irya.unam.mx}}
\affil{Harvard-Smithsonian Center for Astrophysics, Cambridge, Massachusetts, USA  \email{reid@cfa.harvard.edu}}
}

\paperauthor{Laurent Loinard}{l.loinard@irya.unam.mx}{0000-0002-5635-3345}{Centro de Radioastronomia y Astrofisica}{Universidad Nacional Autonoma de Mexico}{Morelia}{Mich}{58089}{Mexico}
\paperauthor{Mark J. Reid}{reid@cfa.harvard.edu}{}{Harvard-Smithsonian Center for Astrophysics}{}{Cambridge}{MA}{02138}{USA}

\begin{abstract}
Various sign-posts of recent star-formation activity, such as water and methanol maser emission or magnetically active low-mass young stars, can be detected with Very Long Baseline Interferometry (VLBI) radio arrays. The extremely accurate astrometry already attainable with VLBI instruments implies that the trigonometric parallax and the proper motion of these objects can be measured to better than 1\% for sources within a few hundred parsecs of the Sun, and better than 10\% for objects at a few kiloparsecs. An ngVLA with baselines extending to several thousand km will have a sensitivity more than one order of magnitude better than current VLBI instruments, and will enable such highly accurate astrometric measurements to be performed throughout the Milky Way. This will provide a full six-dimensional view (three spatial and three velocity coordinates) of star-formation in the Galactic disk, and have a transformative impact on our understanding of both star-formation processes and Galactic structure. 
\end{abstract}

\section{Introduction \label{sect:intro}}

Even when considering only the Milky Way galaxy, star-formation involves a wide variety of relevant spatial scales. Individual protostars accrete through disks that are 10--100 AU across, from protostellar envelopes whose radii are typically 10,000 AU. In addition, stars typically do not form in isolation. Instead, star-formation involves molecular cores, filaments, clouds, and complexes with scales between a tenth and several hundred parsecs. Finally, star-formation in the Galactic disk is organized into a coherent spiral pattern, where individual spiral arm segments can be traced out over tens of kiloparsecs. On all of these scales, many questions regarding the physical processes underlying star-formation remain unanswered. Let us provide just three examples, involving widely different spatial scales. 

\begin{itemize}
\item While a paradigm exists for the formation of isolated low-mass stars (Shu, Adams \& Lizano 1987), the formation of multiple stellar systems remains poorly understood. The two leading (and not necessarily mutually exclusive) contending theories are turbulent and disk fragmentation (Padoan et al.\ 2007; Adams et al.\ 1989). These two theories make different predictions on the distribution of separations in the resulting multiple systems that can be tested through high angular resolution observations (e.g.\ Tobin et al.\ 2016a,b). Tight multiple systems (with sizes of a fraction of an AU) are particularly interesting in this context because their orbital periods are short, so they can be fully characterized within a reasonable amount of time. 

\item Parsec-scale filamentary structures are now known to be ubiquitous in regions of star-formation (Andr\'e et al.\ 2017), but their exact impact on the star-formation process remains unclear. It has been argued that their fragmentation might lead to the formation of dense cores and that gas motion along the filaments could be important to sustain the accretion onto the protostars contained in these cores (Lu et al.\ 2018). To unambiguously test this possibility, it is important to establish the exact orientation of the filaments, the kinematics of the gas motion, as well as the location and kinematics of the young stars embedded within the filaments.

\item Spiral structure is a very common feature in disk galaxies, and it is well established that star-formation (in spiral galaxies) occurs mainly along the spiral arms. Yet, the very origin of the spiral structure and the exact relationship between spiral structure and star-formation remain elusive (see Dobbs et al.\ 2014 for a recent discussion). A popular scenario, the spiral density wave theory of Lin \& Shu (1964) combined with an analysis of the gas response to the resulting stellar potential (Roberts 1969), leads to the prediction of organized {\it streaming motions} across the spiral arms that should be detectable for both the gaseous and the stellar components. However, detailed numerical simulations do not appear to support this theory. Thus, the unambiguous detection of streaming motions (or stringent upper limits on their non existence) would place important constraints on the mechanisms relating spiral structure and star-formation. The necessary observations would have to constrain the exact location and structure of the spiral arms as well as the kinematics of the young stars and the associated ISM. 
\end{itemize}

These three examples illustrate the need for both accurate distances to young stars (to establish the orientation of filaments in star-forming regions or the location of spiral arm features in the Milky Way) and accurate measurements of their 3-dimensional velocity vectors, at scales ranging from individual stellar systems, to specific star-forming regions, to the entire Galactic disk. Formally, this means constraining the distribution function $f(\mathbf{x},\mathbf{\dot{x}})$ for a sufficiently large sample of young stars, and can be achieved through a combination of dedicated spectroscopic and astrometric observations. The spectroscopic observations are necessary to provide the velocity component projected along the line of sight and typically require sensitive optical or infrared observations. The remaining 5 dimensions (3 spatial coordinates, and two velocity components) require accurate positions, trigonometric parallaxes, and proper motions. The Gaia mission will be transformative in providing this type of information for many different populations of astronomical objects, including some young stars. However, given the dust opacity associated with individual star-forming regions and the Galactic plane as a whole, Gaia will be limited to young stars that are both nearby and not deeply enshrouded in dust. Very Long Baseline Interferometry (VLBI) observations at radio wavelengths are immune to dust obscuration and can already deliver trigonometric parallaxes with an accuracy as good as 10 $\mu$as, and proper motions accurate to better than a few $\mu$as yr$^{-1}$ (Reid \& Honma 2014). Taking advantage of this, VLBI observations have been used to study star-forming regions in the Milky Way on a variety of scales\footnote{Other applications include extragalactic proper motions, as well as astrometry of pulsars, compact objects, and evolved stars but they will not be discussed here. See Reid \& Honma (2014) for a review.} as we will describe in Section 2. We will then proceed, in Section 3 and 4, to discuss the impact that baselines in thousands of km on the ngVLA would have on the study of Galactic star-formation, as well the synergies with other existing and planned facilities.

\articlefigure{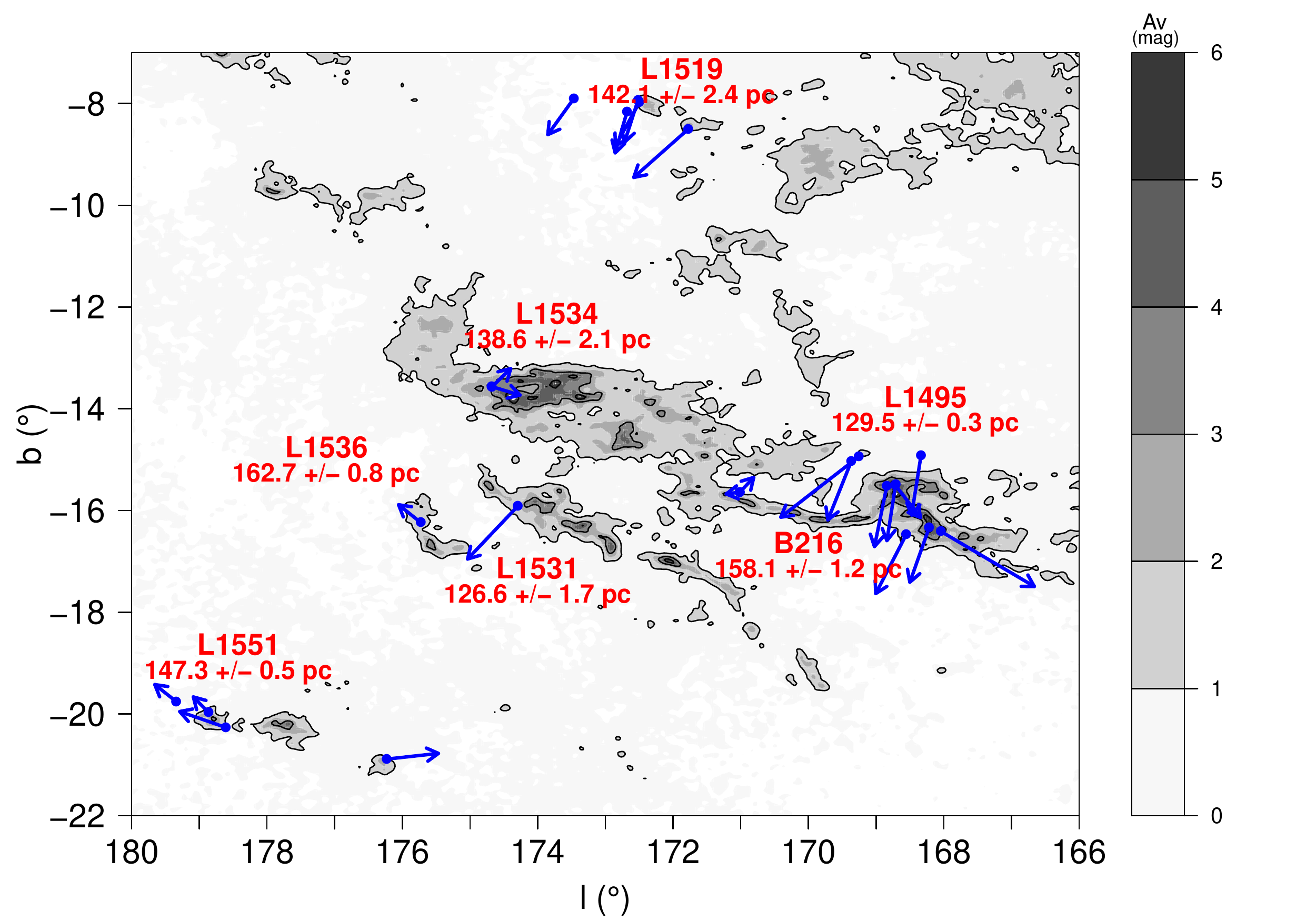}{fig:Taurus}
{Three-dimensional structure of the Taurus star-forming regions (adapted from Galli et al.\ 2018). The underlying grey-scale image shows the extinction map derived by Dobashi et al.\ (2005). The different dust clouds in the region are indicated, together with their distance measured though VLBI astrometric observations. The blue arrows correspond to the proper motions (also determined from VLBI observations) expressed in the LSR. }

\section{Existing VLBI observations \label{sect:already}}

\subsection{Structure and kinematics of nearby regions}

At the distance of the nearest star-forming regions (Taurus, Ophiuchus, Orion, etc.; 100 pc $< d < 500$ pc), VLBI observations of numerous magnetically-active young stars have enabled the measurement of their distance and velocity to an accuracy of about 1\%. For instance, the distance to the Orion Nebula Cluster (ONC; the nearest region of high-mass star-formation) has been refined to 388 $\pm$ 5 pc through a series of VLBI measurements of an increasingly representative sample of individual young stars (Sandstrom et al.\ 2007; Hirota et al.\ 2007; Menten et al.\ 2007; Kim et al.\ 2008; Kounkel et al.\ 2017). An analysis of the Gaia DR2 results by Kounkel et al.\ (2018) on a different subset of young stars in the ONC yields a fully consistent distance determination of 389 $\pm$ 3 pc.

\articlefigure{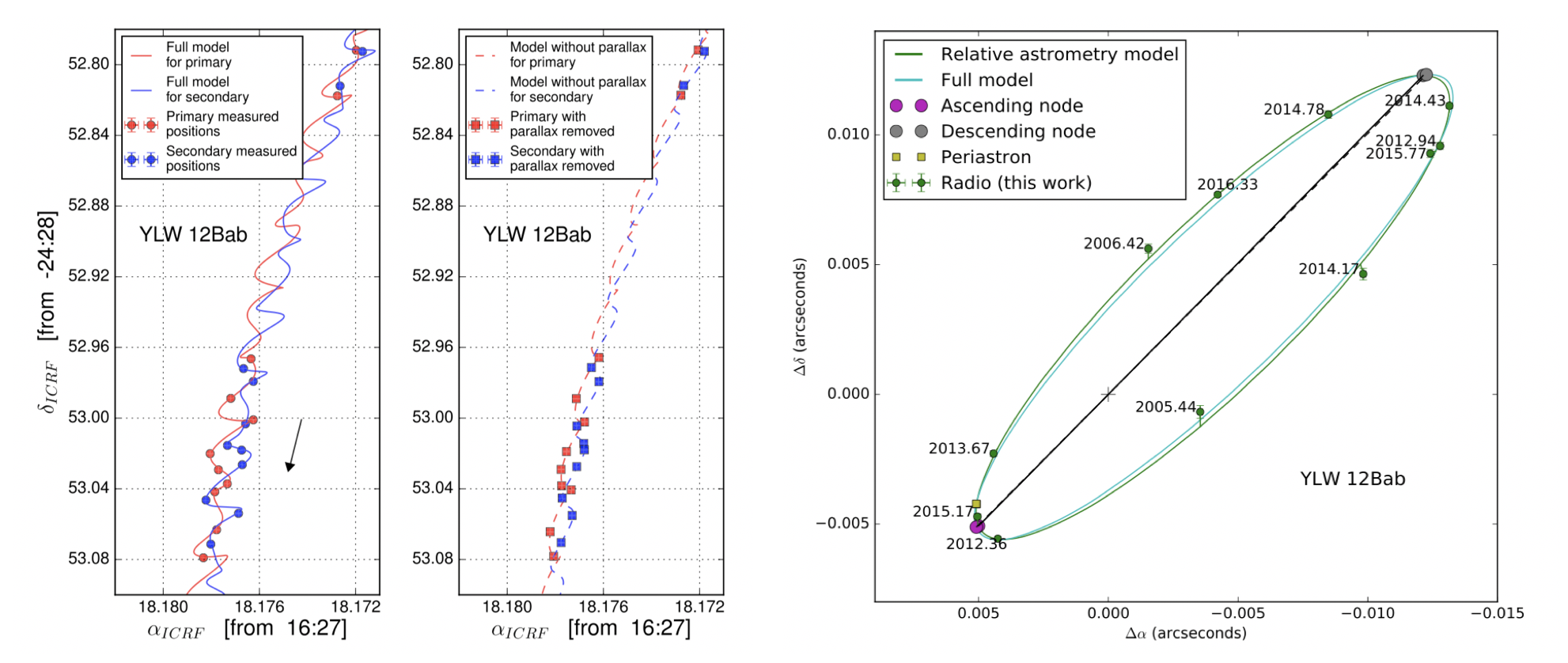}{fig:YLW12B}
{VLBI resolved observations of the tight binary system YLW12B in Ophiuchus (from Ortiz-Le\'on et al.\ 2017a). The leftmost panel shows the positions of the two stars in the systems (as blue and red dots, respectively) measured a different epochs, together with fits to these positions that combine trigonometric parallaxes, systemic, and orbital proper motions. The central panel shows the same positions and fits but with the parallax contribution subtracted. The rightmost panel shows the relative orbit between the two stars in the system with the best Keplerian fit.}

Nearby star-forming regions are typically a few to several tens of parsecs across, so an accuracy of order 1\% on the distance to individual stars is largely sufficient to resolve their depth along the line of sight. As a consequence, VLBI observations have been able to reconstruct the 3D spatial structure of several of these regions. The example of Taurus, where the depth ($\sim$30 pc) is comparable to the extent of the complex on the plane of the sky, is illustrated in Figure 1. A combination of VLBI astrometry and molecular gas spectroscopy can then be used to reconstruct the six-dimensional (space-velocity) structure of the complexes. Preliminary result of this type have been presented by Kounkel et al.\ (2017) who showed that two sub-regions along the Orion A complex (the ONC and Lynds 1641) are approaching each other both in the plane of the sky (as indicated by their relative proper motions) and along the line of sight. The latter conclusion stems from the combined facts that (i) the ONC (at 389 pc) is significantly nearer than L 1641 (at 430 pc), and (ii) the radial velocity of the molecular gas associated with the ONC is less blue-shifted than that of L 1641. 

\articlefigure{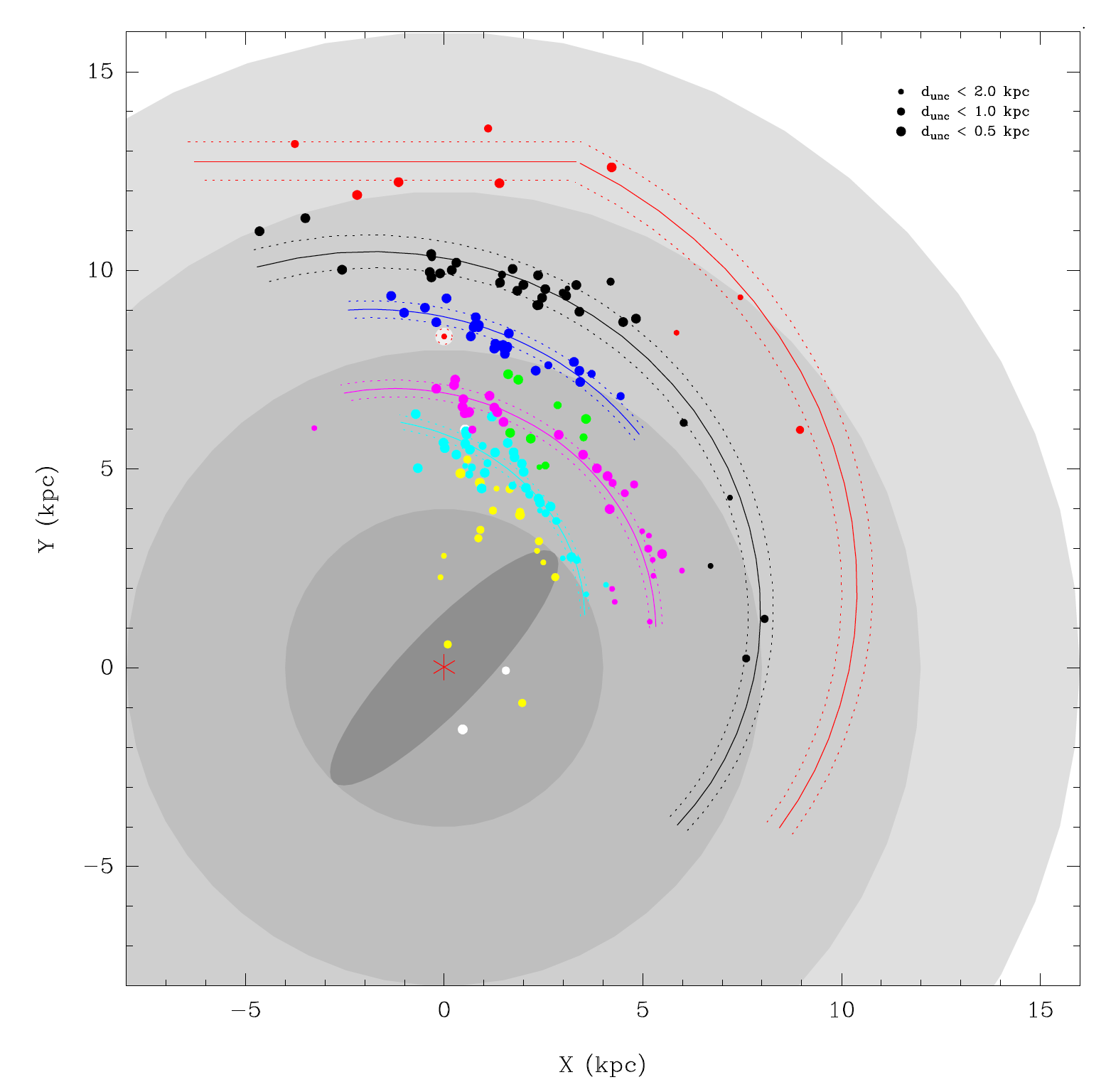}{fig:galaxy}
{Location of high-mass star-forming regions in the Milky way with VLBI-measured distances (from Reid et al.\ 2018, in prep.) overlaid on a face-on sketch of the Galaxy. Different symbols and colors show different arm segments.}

\subsection{Multiple stellar systems}

An interesting by-product of the intensive VLBI observations that led to the accurate distances to nearby star-forming regions mentioned above, has been the identification and spatial resolution of a significant number of very tight binary systems (Ortiz-Le\'on et al.\ 2017a,b, Kounkel et al.\ 2017). As an illustration, Figure 2 shows the case of the YLW12B system in Ophiuchus, which is resolved into two stars in an orbit with a semi-major axis of 12.6 mas (1.73 AU) and an orbital period of 1.42 yr. This tight binary is itself orbited by a third star on a wider orbit, making the entire system a hierarchical triple. Modelling the measured positions with a combination of parallax, systemic, and orbital motions results in extremely accurate mass determinations for the individual stars (1.397 $\pm$ 0.019 \Msun\ and 1.258 $\pm$ 0.006 \Msun). Such observations, obtained for a statistical sample of young multiple systems, provide important constraints for theoretical models of early stellar evolution, and inform models of multiple star-formation on the observed distribution of system architectures. 

It is important to appreciate that two distinct properties of VLBI observations came into play to enable these results. On the one hand, VLBI observations provide coordinates for the individual stars relative to a fixed quasar --and not just relative to one another. This registration of the positions relative to a fixed reference is what enables individual masses to be measured, rather than only the total mass of the system. Second, the very high angular resolution of VLBI images (of order 1 mas) enabled the resolution of the different stars in the systems. This combination of VLBI image properties are not available at any other wavelengths. Stellar mass determination based on orbital motion modelling has been obtained for increasingly tight systems from optical or near-infrared ground-based observations using techniques like adaptive optics and aperture masking. Those observations, however, only provide the relative positions of the stars in a given system so they only yield the total mass of the systems. Individual stellar masses then have to be estimated using model-dependent assumptions on the mass ratio (which tend to be derived from the very theoretical models that the observations intend to constrain). Dedicated astrometric spaced-based missions such as Gaia, on the other hand, do provide absolute coordinates, but cannot resolve tight systems (the angular resolution of Gaia is several tenths of an arcsecond). 

\subsection{Galatic structure}

Using VLBI observations of water and methanol maser emission associated with regions of high-mass star-formation, it is possible to measure their trigonometric parallax to high accuracy. Reid et al.\ (2009, 2014) and Honma et al.\ (2012) have obtained distances to more than 100 regions distributed along Galactic spiral segments within about 10 kpc of the Sun (Figure 3). These results have enabled a huge refinement in the determination of the spiral structure of our Galaxy, and of the fundamental Galactic parameters, such as a distance from the Sun to the Galactic center $R_0$ and the circular velocity at the position of the Sun $\theta_0$. An interesting conclusion of this work is that essentially all high-mass star-formation appear to be associated with know spiral features --i.e.\ there appears to be little or no inter-arm high-mass star-formation activity in our Galaxy. In a remarkable recent result, Sanna et al.\ (2017) were able to measure the distance to a high-mass star-forming region (G007.47+00.05) located on the far side of the Galaxy at more than 20 kpc using VLBI observations of water maser emission. This result is accurate to about 15\%, and demonstrates the potential of very long baseline observations to map the spiral structure over the entire Galactic disk. 

\section{The impact of the ngVLA}  

The examples shown in Section 2 demonstrate that VLBI astrometry has already provided fundamental constraints on star-formation processes at scales ranging from a fraction of an AU to tens of kiloparsecs. Continental-size baselines on the ngVLA will result in a sensitivity more than one order of magnitude better than that of existing VLBI facilities such as the VLBA or the EVN, dramatically increasing the number of detectable sources, both as targets and calibrator. At frequencies above $\sim$ 5 GHz, the astrometric accuracy of VLBI observations for a given maximum baseline length is limited by residual delay errors which scale linearly with the separation between the target and the calibrator. The densification of calibrators resulting from the increased sensitivity of the ngVLA will therefore directly translate to an increase in astrometric accuracy, which will become as good as 1 $\mu$as for trigonometric parallax measurements (Reid et al.\ 2018 in this volume). In other words, {\bf the astrometry delivered by the ngVLA equipped with baselines of thousands of km will surpass that of Gaia by one order of magnitude}. The combination of this and the detectability of weaker target sources will have several very important consequences. 

\begin{itemize}

\item In nearby regions, the ngVLA will deliver distance and velocity measurements accurate to better than 0.1\%. In addition, the number of detectable targets will increase very significantly. At the moment, only about 10--20\% of young stars in a given region is detectable with VLBI arrays. Given the radio luminosity function of young stars (Ortiz-Le\'on et al. 2017a), the ngVLA will be able to detect the majority of sources. This will provide a full six-dimensional description of each individual region, that will be directly comparable with numerical simulation of the formation and evolution of stellar clusters. It will also increase the number of multiple stellar systems for which the architecture and extremely accurate individual masses will become available. As explained in Section 2.2, VLBI observation is currently the only technique that can provide the combination of resolved images and coordinates registered to a fixed reference necessary to measure individual masses for systems tighter than several tenths of an arcsecond.

\item The magnetically active young stars mentioned in Section 2.1 are currently only detectable up to distances of about 500 pc. With the ngVLA, they will be detectable in the nearest spiral arm segments. Since such active stars are typically low-mass, they are much more numerous than the maser sources associated with high-mass protostars. As a consequence, their detection in nearby spiral arm segments will enormously improve the way that spiral arms can traced. This will make it possible to map in detail the distribution of star-formation activity and the kinematics of the associated young stars along and across spiral arm segments. Combined with observations of the molecular ISM component, this will provide fundamental constraints on the star-formation mechanisms in spiral galaxies and their relationship with spiral structure.

\item The ngVLA will be able to detect and measure accurate distance to maser associated with high-mass star-forming regions throughout the Milky Way Galaxy. The measurement reported by Sanna et al.\ (2017) of a parallax for a water maser at more than 20 kpc was only possible because of the extraordinary brightness of that specific maser. With the ngVLA, however, such measurements will become possible for a large population of masers on the far side of the Milky Way. This will permit the delineation of the entire spiral structure of the Milky Way, with fundamental implications for the analysis of star-formation on the scale of the entire Galaxy. 

\end{itemize}

In addition to the purely astrometric results mentioned here, long baselines on the ngVLA will have the potential to reveal interesting physics related to stellar magnetic activity. For instance there is evidence that tight binary systems are more frequently radio-active than single stars (or wider binaries) and that the exact location of the radio emission centroid is modulated by the orbital motion of the system (e.g.\ Massi et al.\ 2002; Torres et al. 2012). The increased sensitivity and astrometric accuracy of the ngVLA will provide statistically relevant constraints on these and a variety of other phenomena that might prove relevant to exo-space wheather.

\section{Synergies at other wavelengths and with other instruments}

As detailed above, the ngVLA equipped with very long baselines will provide a tomographic six-dimensional view of star-formation in our Galaxy, providing information on scales from 1 AU to tens of kpc. This will be highly complementary of the Gaia satellite to reveal the dust obscured Galactic plane and dust-enshrouded nearby star-formation. In addition, since many ingredients enter the process of star-formation, observations at multiple wavelengths are necessary to make progress. The ngVLA itself, particularly combined with ALMA, large single dish telescopes, and future infrared space observatories, will enable the mapping of dust, molecular, atomic, and ionized gas at high spatial resolution in individual protostars and extended star-forming regions. Such observations might also reveal the role of magnetic fields both at large and small scales. The study of the physical mechanisms leading to continuum radio emission in low-mass young stars and their relationship to binarity and other stellar properties will require low frequency observations from the ngVLA itself, but also from the SKA. Finally, X-ray telescopes will be useful both to probe ISM distributions and stellar activity. 

\acknowledgements L.L.\ acknowledges the financial support of DGAPA, UNAM (project IN112417), and CONACyT, M\'exico.

\end{document}